\documentclass[aps,prl,showpacs,superscriptaddress,twocolumn,10pt,floatfix]{revtex4-1}
\usepackage{graphicx}
\usepackage{amsmath, amsthm, amssymb}
\usepackage{bm}
\usepackage{microtype}
\usepackage{color}
\usepackage{hyperref}

\newcommand{\mb}[1]{\mathbf{#1}}
\newcommand{\pd}{\partial}
\newcommand{\tr}{\textrm{Tr}}
\newcommand{\up}{\uparrow}
\newcommand{\down}{\downarrow}

\newcommand{\mc}[1]{\mathcal{#1}}
\newcommand{\hc}{{\rm H.c.}}
\newcommand{\avg}[1]{\langle #1\rangle}

\def\e{\epsilon}
\def\la{\langle}
\def\ra{\rangle}

\begin{document}
\title{Dynamical response of dissipative helical edge states}
\author{Doru Sticlet}
\email{doru-cristian.sticlet@u-bordeaux.fr}
\affiliation{LOMA (UMR-5798), CNRS and University Bordeaux 1, F-33045 Talence, France}
\author{J\'er\^ome Cayssol}
\email{jerome.cayssol@u-bordeaux.fr}
\affiliation{LOMA (UMR-5798), CNRS and University Bordeaux 1, F-33045 Talence, France}
\begin{abstract}
Quantum spin Hall insulators are characterized by topologically protected counterpropagating edge states. Here we study the dynamical response of these helical edge states under a time-dependent flux biasing, in the presence of a heat bath. It is shown that the relaxation time of the edge carriers can be determined from a measurement of the dissipative response of topological insulator disks. The effects of various perturbations, including Zeeman coupling and disorder, are also discussed. 
\end{abstract}
\maketitle

{\itshape Introduction.---}The hallmark of two-dimensional (2D) quantum spin Hall (QSH) topological insulators (TIs) consists in the existence of dissipationless conducting edge states in the absence of any time-reversal breaking perturbations~\cite{Hasan2010,*Qi2011}. Due to spin-orbit coupling and a particular bulk band structure, the edge carriers' spin is tied to their momentum~\cite{Kane2005,Bernevig2006}. These helical edge states have been reported experimentally in HgTe/CdTe~\cite{Koenig2007,*Roth2009} and InAs/GaSb~\cite{Knez2011} quantum wells.
So far, most of the studies have covered the equilibrium or ground-state physics of helical edge states, while less is known about their dynamics and the associated relaxation mechanisms.
Only recently, the problem of dissipation has gained attention in the context of topological insulators (TI)~\cite{Viyuela2012,*Rivas2013} and topological superconductors~\cite{Diehl2011,*Bardyn2013}.

Recently, it has been proposed that the Floquet type of TIs can be engineered by applying a proper external drive on semimetals or trivial band insulators~\cite{Oka2009,*Lindner2011,*Kitagawa2011, *Cayssol2013,*Rudner2013}. Floquet bands have already been reported in time-resolved photoemission experiments on three-dimensional TIs~\cite{Wang2013}, and their topological nature is under active debate. Relaxation phenomena are crucial to establish such nonequilibrium steady states of matter, and ensure the balance between the energy injected by the drive and the energy dissipated towards microscopic degrees of freedom of the environment.

Meanwhile, experimental progress has been achieved in extracting typical relaxation times of carriers in coherent conductors such as normal-superconducting (NS) rings~\cite{Chiodi2011, Dassonneville2013}. The idea is to couple a small coherent system, characterized by a flux-dependent spectrum, to a multimode superconducting resonator. The dissipative and nondissipative magnetic susceptibility of unconnected samples is obtained by measuring the energy shifts and quality factors of the resonances as a function of frequency, temperature, and dc magnetic flux. In this Rapid Communication, we suggest that these techniques could be applied to extract the typical relaxation times of helical edge carriers circulating around disks of two-dimensional (2D) TIs.

In view of these experimental advances, this Rapid Comm. addresses the dynamical response of the generic helical edge state of a 2D TI coupled to a thermal bath and threaded by a time-dependent flux $\Phi(t)$, which is the superposition of a dc flux $\phi$ and a small alternating flux at a single frequency $\omega$ (see experiments~\cite{Chiodi2011, Dassonneville2013}).
It is obtained that the dissipative response of the helical edge state exhibits a characteristic phase-dependent signature: a single peak is located either at $\phi=0$ or at $\phi=\phi_0/2$, depending on the electronic filling.
This peak has a maximal amplitude when the frequency is equal to the relaxation rate of the edge carriers. In contrast to standard metallic rings~\cite{Trivedi1988,Reulet1994} or NS rings~\cite{Chiodi2011, Dassonneville2013}, the extraction of the carrier lifetime is simplified by a selection rule which forbids interband transitions between left and right spin-polarized movers. This is a dynamical manifestation of the edge states' helical structure.
These analytical results are validated in a comparison with lattice simulations of the Bernevig-Hughes-Zhang (BHZ) model for HgCd/CdTe quantum wells~\cite{Bernevig2006}. Lastly, this Rapid Communication analyzes the effects of a Zeeman spin-flip coupling and of disorder on the predicted phenomena.

\begin{figure}[t]
\includegraphics[width=0.65\columnwidth]{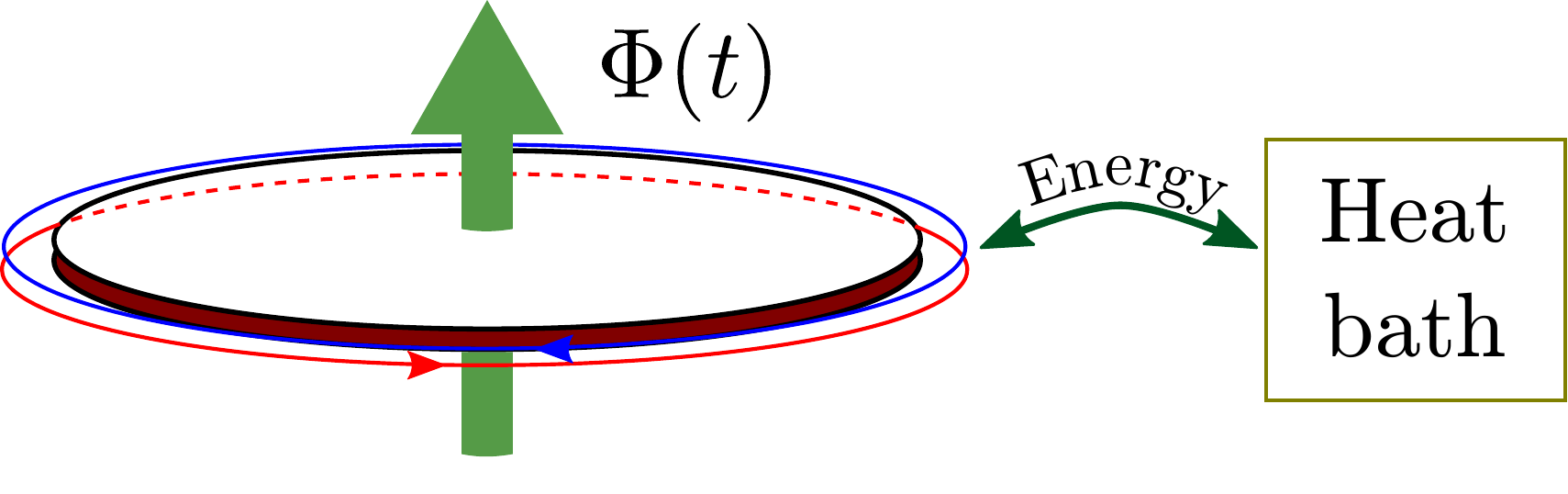}
\caption{(Color online) A QSH TI disk under a time-dependent perpendicular magnetic field $B(t)$. The Hamiltonian~(\ref{hel}) models the counterpropagating edge states (red and blue) which enclose a flux $\Phi(t)=\phi+\delta\phi(t)$. The dc flux $\phi$ is varied arbitrarily, while the time-dependent oscillatory flux $\delta\phi(t)=\delta\phi_{\omega}\cos\omega t$ has a small amplitude with respect to the flux quantum. The edge carriers are exchanging energy with a heat bath and the dynamical susceptibility $\chi(\omega)$ gains a dissipative component at finite frequency.}
\label{fig:model}
\end{figure}

{\itshape Model and formalism.---}Let us consider a disk of a 2D TI under a perpendicular time-dependent uniform magnetic field $B(t)$ (Fig.~\ref{fig:model}). Here, the focus is on the response of the helical edge liquid which encloses the time-dependent magnetic flux $\Phi(t)=\phi + \delta\phi(t)$, $\phi$ being a constant flux and $\delta\phi(t)=\delta\phi_{\omega} \cos\omega t$ being a small oscillating flux.
The ac amplitude $\delta\phi_{\omega}$ is kept much smaller than the magnetic flux quantum $\phi_0=h/e$, $h$ being the Planck constant and $e$ the absolute value of the electronic charge.
The total Hamiltonian $H$ describing the system decomposes into a static and a dynamic part as $H=H_0+H'(t)$, with
\begin{equation}\label{hel}
H_0=\frac{h v_F}{L}\bigg(-i\frac{\pd}{\pd\theta}+\frac{\phi}{\phi_0}
\bigg)\sigma_3,\quad H'(t)=\frac{ev_F}{L}\delta\phi(t)\sigma_3.
\end{equation} 
The Fermi velocity of the carriers is $v_F$, the length of the edge state, $L$, and the angular coordinate, $\theta$. The $\sigma_3$ matrix is the standard diagonal spin Pauli matrix.

In the absence of a time-dependent drive [$\delta\phi(t)=0$], the helical liquid is described by the low-energy effective Hamiltonian $H_0$, and it supports a robust persistent current $I_{\rm per}(\phi)$, characterized by a maximal amplitude $I_0 = ev_F /L$ at zero temperature~\cite{Sticlet2013}. The flux-dependent energy levels $\epsilon_{n \sigma}(\phi)=\e_{\mb n}(\phi)=\sigma \hbar \omega_0(n+\phi/\phi_0)$
are discrete and identified by an angular momentum $n$ and a spin $\sigma$ quantum numbers, which are gathered in the notation $\mb n=(n,\sigma)$. The corresponding energy eigenstates solve the Schr\"odinger equation $H_0|\mb n\ra=\e_{\mb n}(\phi)|\mb n\ra$, where $|\mb n\ra$ are the eigenspinors of $\sigma_3$ times $e^{in\theta}$. The energy spacing between adjacent levels of a given spin and flux is denoted by $\hbar\omega_0=h v_F/L$. Each energy level carries a flux-independent current $i_{\mb n}=-\sigma I_0=-\sigma ev_F/L$.

Let us consider that the quantum edge states are coupled to a thermal bath containing many degrees of freedom. These degrees of freedom could have various distinct microscopic origins: electromagnetic modes of the external circuit, phonons, bulk states of the disk, etc. Then, in response to the finite driving term $\delta\phi(t)=\delta\phi_{\omega} \cos \omega t$, the edge supports both nondissipative ($I_{\rm ac}' \cos \omega t $) and dissipative ($I_{\rm ac}'' \sin \omega t$) ac steady currents. This response is captured by a complex frequency-dependent susceptibility $\chi(\omega)=\chi'(\omega)+i\chi''(\omega)$, defined by $\chi'(\omega)=I_{\omega}'/\delta\phi_{\omega}$ and $\chi''(\omega)=I_{\rm ac}''/\delta\phi_{\omega}$.
In the present setup, the TI disk is unconnected and therefore it only exchanges energy with the environment, while the number of particles remains fixed.

Here, we will not investigate the microscopic mechanisms leading to dissipation, but rather provide a generic and phenomenological model to describe it in the case of a weak coupling to the environment. To this aim, we consider the evolution of the system under the following kinetic equation for the reduced (single-particle) density operator $\rho(t)$ (obtained after tracing out the environment degrees of freedom~\cite{Trivedi1988,Reulet1994,Larkin1986, *Browne1987,*Weiss2008}):
\begin{equation}\label{master}
\frac{\pd \rho(t)}{\pd t}+\frac{i}{\hbar}[H(t),\rho(t)]
=-\gamma[\rho(t)-\rho_{\rm qe}(t)],
\end{equation}
where $\rho_{\rm qe}(t)=\{\exp[(H(t)-\mu)/k_B T]+1\}^{-1}$ is the quasi-equilibrium density matrix at temperature $T$, $k_B$ being the Boltzmann constant. The matrix $\gamma$ phenomenologically represents the relaxation rates for populations and coherences in the density matrix operator. Because the system exchanges only heat with the environment, the number of particles is fixed. Consequently, the chemical potential $\mu$ is not constant and generally depends on flux, number of particles, temperature, and time. Nevertheless, $\mu$ can be taken here constant in flux, due to the particular flux dependence of the last occupied energy level for a given parity of electron number. Moreover, the time dependence of $\mu$ brings only a negligible contribution to the dissipative response in comparison with other competing terms~\cite{SM}.

In the linear response approximation ($\delta\phi_\omega \ll \phi_0$), the master equation~(\ref{master}) is solvable, and the complex linear susceptibility can be decomposed into three parts~\cite{Trivedi1988,Reulet1994},
\begin{equation}\label{chiPart}
\chi(\omega,\phi)=\chi_{\rm per}
+\chi_D(\omega,\phi)+\chi_{ND}(\omega,\phi).
\end{equation}
The static part of the susceptibility $\chi_{\rm per}$ is purely real and it is due to the persistent current in the system. The second and third terms are called diagonal and nondiagonal with reference to the $H_0$ eigenstate basis. The diagonal susceptibility $\chi_D$ describes only the intraband response of the system, while the nondiagonal susceptibility $\chi_{ND}$ is related to interband transitions. Note that all the terms in Eq.~(\ref{chiPart}) depend also on temperature. 

{\itshape Helical edge states' susceptibility.---}
The static part of the susceptibility $\chi_{\rm per}=\frac{\pd}{\pd\phi}(\sum_\mb n i_\mb nf_\mb n)$ is the derivative of the persistent current with respect to the dc flux $\phi$. In this case, the sum runs over the angular momentum and spin quantum numbers. The functions $f_\mb n$ represent henceforth the Fermi-Dirac distribution function for static Hamiltonian $H_0$, $f_\mb n=f(\e_\mb n(\phi))$.

The perturbation $H'(t)$ commutes with $H_0$ and it cannot induce spin flips or changes in the angular momentum of the electrons. Because the system does not exchange electrons with the environment the spin and angular quantum numbers remain conserved. This selection rule forbids interband transitions and it implies that the nondiagonal susceptibility $\chi_{ND}(\omega,T,\phi)$ vanishes. Therefore, dissipation can occur only through intraband relaxation processes. 

\begin{figure}[t]
\centering
\includegraphics[width=0.48\columnwidth]{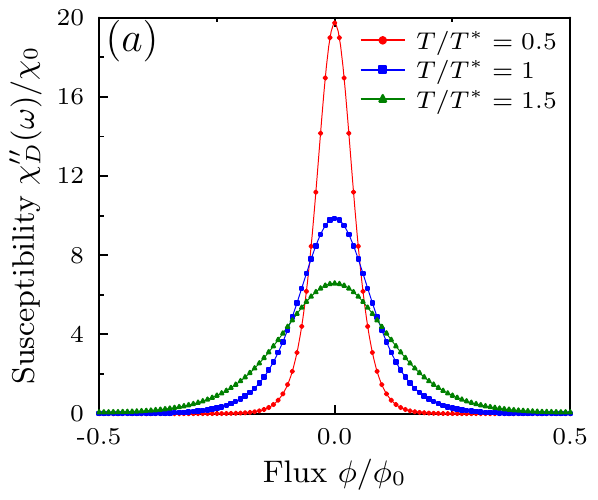}
\includegraphics[width=0.48\columnwidth]{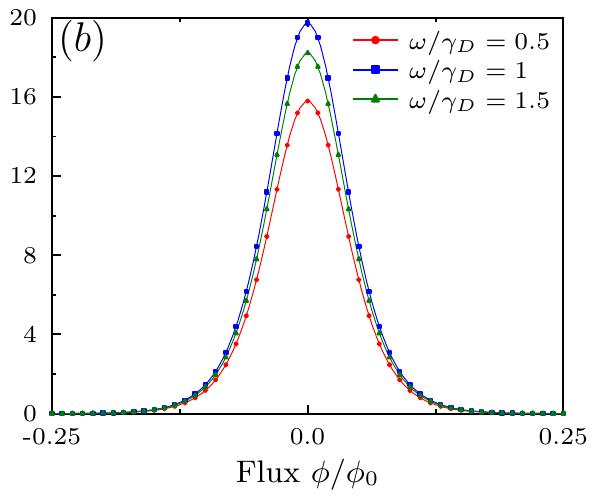}
\includegraphics[width=0.48\columnwidth]{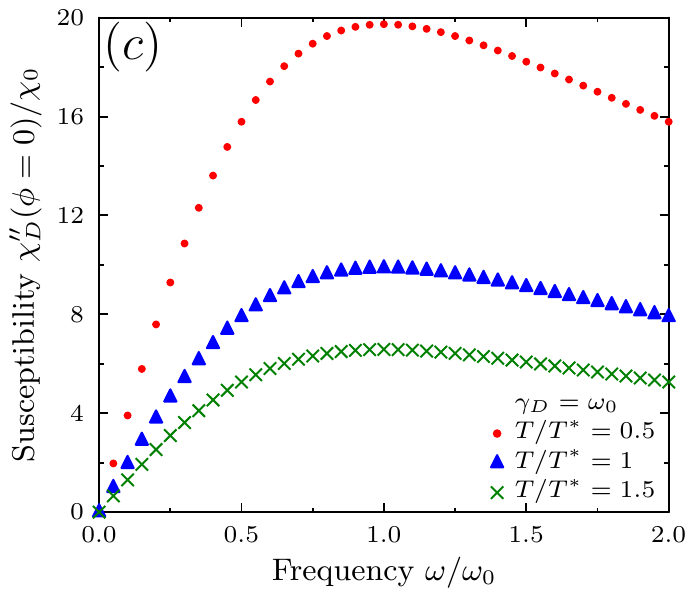}
\includegraphics[width=0.48\columnwidth]{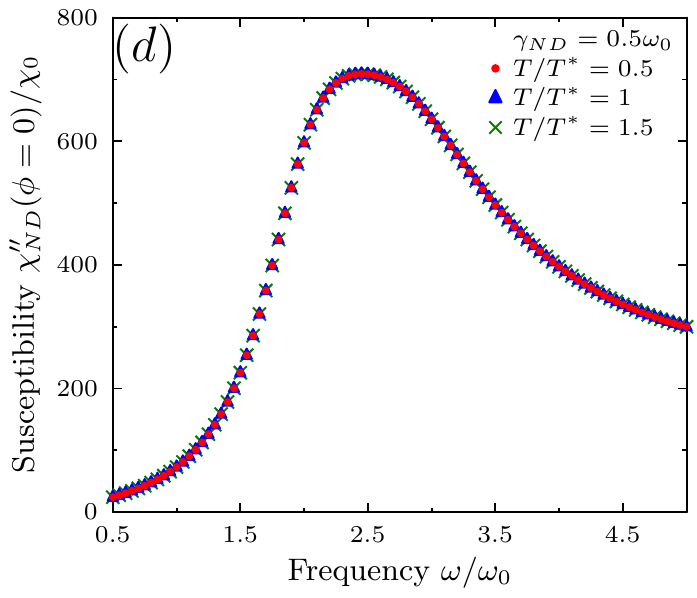}
\caption{(Color online) For the clean helical edge, the dissipative susceptibility $\chi''_D(\phi,\omega)$ has a peak at zero flux $\phi=0$ (a) and is maximal at frequency $\omega=\gamma_D^{-1}$ [b and c]. The analytical result (continuous lines) and the Bernevig-Hughes-Zhang (BHZ) lattice model (markers) coincide (after multiplying by a factor 2 the single helical edge result, in order to take into account the presence of two edges in the lattice simulations). (d) In the lattice model, there is a large nondiagonal contribution $\chi''_{ND}$ which is independent of the temperature. Parameters of the system (unless otherwise stated): temperature $T=0.5T^*$,  frequency $\omega=\omega_0$, and lattice size $(L_x,L_y)=(80,80)a$. In the BHZ model $(A,B,M)=(1,0.6,1)$.}
\label{fig:clean}
\end{figure}

This is a remarkable simplification with respect to the case of multilevels systems encountered in experiments for normal (and Josephson) rings, where separating the three contributions in Eq.~(\ref{chiPart}) is a difficult and subtle task~\cite{Reulet1994, Dassonneville2013, Ferrier2013}. Therefore, the linear susceptibility of the helical edge contains only two terms: $\chi(\omega,\phi)=\chi_{\rm per}+\chi_D(\omega)$.
Furthermore, the dissipative part of the susceptibility has only one term, $\chi''(\omega)=\chi''_D(\omega)$, since $\chi_{\rm per}$ is purely real (nondissipative).
Moreover, the diagonal rates $\gamma_\mb{nn}$ are assumed to be all identical $\gamma_\mb{nn}=\gamma_D$ and flux independent, since the energy levels
are equidistant and have the same absolute value of the level current.
The dissipative response $\chi''_D$ is given by the imaginary part of the diagonal susceptibility~\cite{SM},
\begin{equation}
\chi_{D}(\omega,\phi)= \frac{\gamma_D}{i\omega-\gamma_D} \sum_\mb n i_\mb n^2
\frac{\pd f_\mb n}{\pd\e_\mb n},
\end{equation}
and it is maximal for $\omega=\gamma_D$ (Fig.~\ref{fig:clean}). The edge states' lifetime $\gamma_D^{-1}$ can then be measured from dissipative response by sweeping the driving frequency.

The explicit result for dissipative susceptibility as a function of temperature, flux, frequency, and chemical potential $\mu$ reads as~\cite{SM}
\begin{eqnarray}\label{chidiss}
\frac{\chi_D''(\omega,\phi)}{\chi_0}&=&\frac{4\omega\gamma_D}{\omega^2+\gamma_D^2}
\bigg[\frac{1}{2}+\sum_{m=1}^\infty\frac{mT/T^*}{\sinh(mT/T^*)}\notag\\
&&\times\cos(2\pi m\frac{\phi}{\phi_0})
\cos(2\pi m\frac{\mu}{\hbar\omega_0})\bigg],
\end{eqnarray}
in units of $\chi_0=\frac{I_0}{\phi_0}=\frac{e^2v_F}{hL}$. The characteristic temperature $T^*$  is proportional to the level spacing, $T^*=\hbar v_F/(\pi k_B L)$.
It immediately follows that the dissipative susceptibility of the current is peaked at zero flux (Fig.~\ref{fig:clean}). If the fermionic parity is changed by adding or substracting a single particle, the chemical potential changes by $\hbar\omega_0/2$ and the peak moves to half-integer flux $\phi/\phi_0=\pm 0.5$~\cite{SM}.

Since the current matrix is diagonal, there is no damping rate for coherences $\rho_{\mb m\mb n}$ $(\mb m\ne\mb n)$, and the dissipation is entirely captured by the evolution for the populations $\rho_{\mb n\mb n}$ as in ~\cite{Landauer1985,*Buettikker1986}. The eventual contribution of bulk states in the insulating disk will be discussed below in connection with the lattice simulations.

{\itshape Comparison with the BHZ model.---} We now present numerical simulations supporting the analytical results above. We use the Bernevig-Hughes-Zhang (BHZ) model on a square lattice, described by the Hamiltonian~\cite{Bernevig2006}
\begin{eqnarray}\label{bhz}
H&&=\mathop{\sum\sum}_{x=1\,y=1}^{L_x\,L_y-1}
c^\dag_{xy}\big[\big(\frac{A}{2i}\sigma_1\tau_3+B\sigma_3\tau_0 \big)e^{i\frac{\varphi(t)}{L_x}}c_{x+1y}\\
&&+\big(\frac{A}{2i}\sigma_2+B\sigma_3 \big)\tau_0c_{xy+1}
+\big( \frac{M}{2}-2B \big) \sigma_3\tau_0c_{xy}\big]+\hc,\notag
\end{eqnarray}
$A$, $B$, and $M$ being material parameters, and $\varphi(t)=2\pi\Phi(t)/\phi_0$ the time-dependent phase induced by the applied flux (lattice constant $a=1$). This model is a useful lattice regularization of the effective 4-band Dirac model describing the topological transition in HgTe/CdTe quantum wells~\cite{Bernevig2006}. The different terms are tensor products of the Pauli matrices $\sigma$ and $\tau$ describing internal
degrees of freedom.
We use the hollow cylinder geometry, with base circumference $L_x$ and height $L_y$. The system is taken in a topological insulating phase (bulk gap $\simeq 2A$) and at half filling $N_{1/2}=2L_xL_y$. Then the model in Eq.~(\ref{bhz}) exhibits a pair of counterpropagating helical edge states located at the bottom ($y=1$) (and one at the top ($y=L_y$)) base of the cylinder. At low energy, each pair of edge states is described by the helical model in Eq.~(\ref{hel}) with $v_F=aA/\hbar$. The mapping between the two models requires that the temperature $k_BT$ is taken much smaller than the BHZ bulk gap and also that $L_y$ is large enough to avoid overlap between these two edge states.

In Figs.~\ref{fig:clean}(a) and~\ref{fig:clean}(b) the diagonal susceptibility shows a peak at zero flux which is maximal when the frequency is exactly equal to the relaxation rate $\gamma_D$. The magnetic signal from the helical model [Eq.~(\ref{hel})] is scaled by a factor of 2 in order to account for the two helical edge liquids in the lattice BHZ model (top and bottom of the cylinder). The match between the helical and the BHZ models holds at different driving frequencies, temperatures, or diagonal rates $\gamma_D$. Indeed, the diagonal susceptibility depends crucially on the states near the chemical potential and thus at half filling it is well approximated by that of the edge states inhabiting the gap, while the bulk contribution is negligible. 
The two pairs of edge states must be well separated otherwise hybridization of edge states leads to a vanishing zero-flux susceptibility. Furthermore, the diagonal susceptibility in zero flux decreases with temperature, but it maintains a maximum at $\omega=\gamma_D$ [Fig.~\ref{fig:clean}(c)].  If a pair of particles is added, the susceptibility-flux characteristic is shifted by half-integer flux quantum, such that the susceptibility peak moves to $\phi/\phi_0=0.5$. At odd number of particles the peaks are smaller and appear at both $\phi/\phi_0=0$ and 0.5~\cite{SM}. 

In contrast with the 1D helical model Eq.~(\ref{hel}), the 2D lattice model allows transitions between the bulk states. These transitions induce a large contribution only to the nondiagonal susceptibility $\chi''_{ND}(\omega)\propto L_xL_y$  [Fig.~\ref{fig:clean}(d)], which scales with the number of electrons in the system, while the diagonal contribution scales with the edge length, $\chi''_{D}(\omega)\propto L_x$. Nevertheless, this large bulk-states contribution is almost flux independent in the thermodynamic limit, thereby allowing an easy extraction of the flux-dependent edge contribution~\cite{SM} and determination of the lifetime $\gamma_D^{-1}$ of the edge states. Note that the nondiagonal dissipative response $\chi''_{ND}$ has been evaluated under the assumption that all damping rates for coherences [in Eq.~(\ref{master})] are equal and constant in flux or temperature, $\gamma_\mb{mn}=\gamma_{ND}$.

\begin{figure}[t]
\includegraphics[width=0.48\columnwidth]{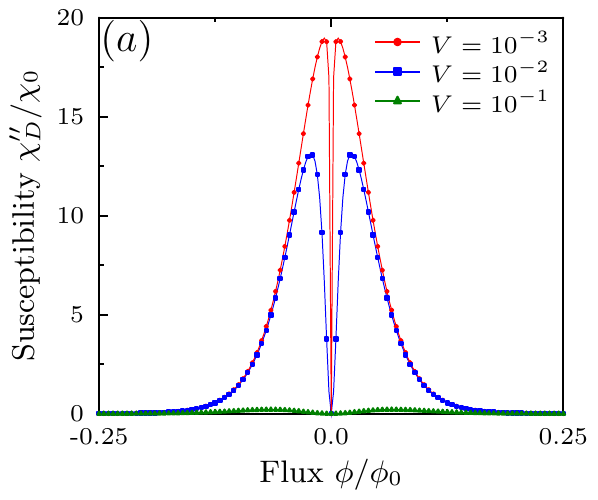}
\includegraphics[width=0.48\columnwidth]{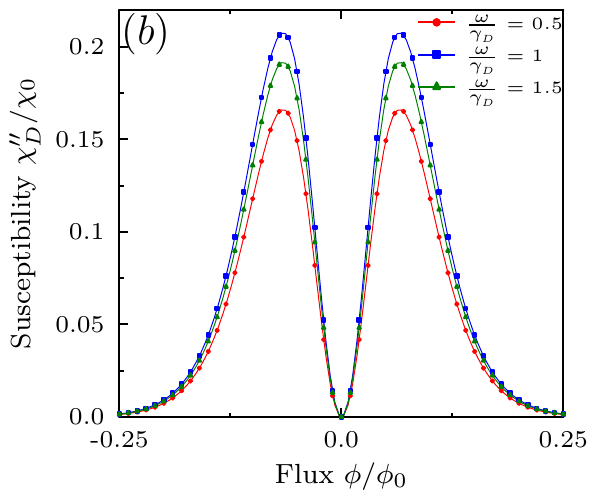}
\includegraphics[width=0.48\columnwidth]{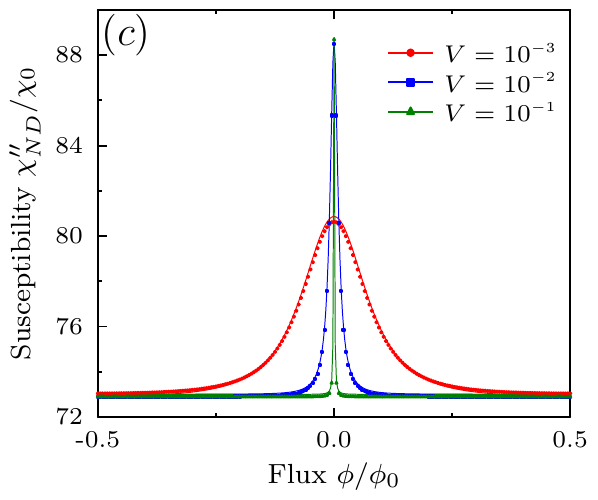}
\includegraphics[width=0.48\columnwidth]{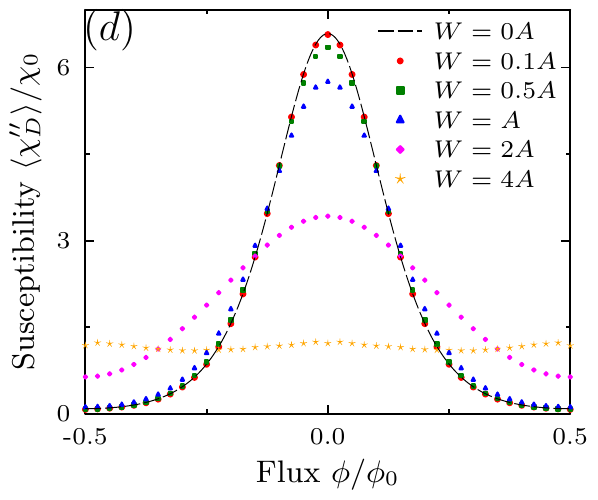}
\caption{
(Color online) The dissipative susceptibility under the effect a small uniform Zeeman field (a), (b), and (c), or under scalar disorder with strength $W$ (d). Both helical model (continuous lines) and BHZ (markers) develop a nondiagonal susceptibility. (a), (b) The diagonal part of the dissipative susceptibility vanishes in zero flux even for a small perturbation $V$. 
(c) The dissipative nondiagonal susceptibility $\chi''_{ND}$ contains a large flux-independent contribution from BHZ bulk states. The response in the gapped helical model, translated by a constant value, matches the lattice result.
The usual Zeeman energy is $V=0.1\hbar\omega_0$. 
Lattice size $(L_x,L_y)=(80,80)a$ and temperature $T=0.5T^*$. (d) The average diagonal susceptibility $\avg{\chi''_D}$ over 200 disorder realizations.
The temperature is $T=1.5T^*$ and lattice size $(L_x,L_y)=(8,60)a$.
The BHZ parameters are $(A,B,M)=(1,0.6.1)$, and the diagonal response is maximal for $\omega/\gamma_D=1$.
}
\label{fig:V}
\end{figure}

{\itshape Disorder effects.---}The addition of scalar disorder does not destroy the edge states. The signature peak in the dissipative diagonal susceptibility slowly decreases; however it does not vanish, if the disorder strength is smaller than the bulk gap [$\simeq 2A$ for the parameters in simulations of Fig.~\ref{fig:V}(d)]. As disorder strength increases and becomes larger than the bulk gap, dips can develop in the diagonal susceptibility in random samples. On average, the susceptibility at large disorder becomes more and more flat and flux independent [Fig.~\ref{fig:V}(d)].

It is important to remark that this situation is different from the case of a regular system with nonrelativistic fermions. Indeed, rings with nonrelativistic fermions present energy level crossings in the ballistic limit which are not protected against disorder: infinitesimal scalar disorder removes the degeneracies and yields a vanishing zero-flux diagonal susceptibility (instead of the peak predicted in the topologically protected edge state).

{\itshape Effect of an in-plane field.---}
An additionally static field induces a Zeeman coupling between spin up and spin down. In the helical model, we consider a constant term proportional to a spin-mixing matrix $\sigma_1$, $H=H_0+H'(t)+V\sigma_1$. In this case the edge states are gapped out in zero flux, leading to a vanishing dissipative diagonal susceptibility. Moreover, the in-gap states now bring a nondiagonal susceptibility $\chi_{ND}$~\cite{SM}. The response will depend on the nondiagonal damping rates $\gamma_\mb{mn}$, which renders the analysis more difficult.

The agreement between the helical and BHZ models still holds [Figs.~\ref{fig:V}(a) and~\ref{fig:V}(b)]. The bulk is largely unaffected by the flux, and its contribution to susceptibility remains almost constant in flux. The features in the nondiagonal susceptibility can be accounted for by the edge state contribution, shifted with a large constant dissipative bulk contribution [Fig.~\ref{fig:V} (c)]. Note that very small spin mixing still opens a gap at time-reversal invariant fluxes. Then the vanishing level current leads to a dip in diagonal susceptibility at zero flux. Only energy states close to these flux values are affected for very small Zeeman fields. These leads to dips in the diagonal susceptibility of small width in comparison to the overall width of the signal. 

Using the material parameters~\cite{Koenig2007} of the HgTe/CdTe quantum wells, we estimate the relevant quantities. The Fermi velocity for HgTe/CdTe quantum wells of thickness $d\simeq 7$~nm is approximately $v_F\simeq 5.5\times 10^5$~m/s. Therefore the characteristic temperature for a ring of size $L=0.5 $~$\mu$m is $T^*\simeq 2.7$~K. The distance between levels at the Fermi surface is $\hbar\omega_0$, which for our given wire sets the characteristic frequency $\omega_0\simeq 6.9\times 10^{12}$~s$^{-1}$. The characteristic current for the same ring length $I_0\simeq 176$~nA. Therefore the characteristic dimensional susceptibility reads $\chi_0\simeq4.26\times 10^7\,\rm H^{-1}$.
In order to explore the physics of the edge states, the temperature was taken smaller than the gap [where the gap $\simeq 2$ for $(A,B,M)=(1,0.6,1)$]. Finally we note that the kinetic equation approach is valid only for temperatures larger than the level broadening, which is the case in all the simulations.

{\itshape Conclusions.---}In this Rapid Communication, we have studied the dissipative response of a 2D QSH insulator under the effect of a small time-dependent driving in flux. Using a helical model for the edge states and exact diagonalization of a tight-binding BHZ insulator, we have proven that the contribution of the edge states and the bulk can be differentiated. Crucially, the lifetime of the edge states can be identified by measuring the frequency where the dissipative response is maximal. While the bulk may 
bring a large contribution to susceptibility, it can be eliminated by observing that it is almost constant in flux.

Moreover, the diagonal dissipative susceptibility is sensitive to the gapping of the edge states (either due to hybridization between pairs of edge states brought in spatial proximity or due to a Zeeman fields at zero flux). The peak in the diagonal susceptibility may split into two (or evem more) peaks into these cases.

\textit{Acknowledgements.}---The authors thank H.~Bouchiat and B.~Dassonneville for introducing them to the topic of dissipation measurements in mesoscopic rings and are grateful to R.~Avriller for careful reading of the manuscript. D.~S. also thanks F.~Pi\'echon and J.-N. Fuchs for stimulating discussions. This work was supported by the French ANR through projects ISOTOP and MASH.
\bibliographystyle{apsrev4-1}
\bibliography{bibl}
\clearpage
\onecolumngrid
\begin{center}
{\bfseries\large Supplemental Material for ``Dynamical response of dissipative helical edge state''}\\
\vspace{1em}
{Doru Sticlet and J\'er\^ome Cayssol}\\
\vspace{0.25em}
{\itshape LOMA (UMR-5798), CNRS and University Bordeaux 1, F-33045 Talence, France}\\
\vspace{0.75em}
\parbox{0.78\textwidth}{\small
The following sections detail the derivation of the results. The first section reviews the general formalism for obtaining the dissipative suceptibility in annular structures under driving. The second section applies the formalism to the helical model and the gapped helical model. Finally, the third section details the comparison between the BHZ and the helical models. It also discusses the dependence of the susceptibility on the number of particles, away from half filling, in the BHZ model.}\\
\end{center}
\twocolumngrid

\section{Ring susceptibility for a small time-dependent perturbation}
This section reviews the linear response theory yielding the total susceptibility for rings threaded by a time-dependent flux, following Ref.~\onlinecite{Trivedi1988}.

Let us consider a static Hamiltonian which depends on flux $H_0(\phi)$ with a time perturbation $H'(t)$ due to an oscillating flux $\delta\phi(t)=\delta\phi_\omega\cos(\omega t)$. 

The unperturbed Hamiltonian $H_0$ represents a lattice system with a discrete number of states. It has a set of energy eigenvalues $\e_n$ and an orthonormal set of eigenvectors $\{|n\ra\}$. The following static operators are used throughout the Supplement: the current $J=-\pd H_0/\pd\phi$ and susceptibility $X=\pd J/\pd\phi$ operators. They are generally not diagonal in the basis of $H_0$. 

The amplitude of the oscillating flux is very small with respect to the flux quantum, $\delta\phi_\omega\ll\phi_0$. In linear response,
any physical quantity is expanded in $\delta\phi_\omega$ and it is sufficient to consider a single Fourier component at the oscillating frequency $\omega$. The total flux is the real part of $\Phi(t)=\phi+\delta\phi_\omega e^{-i\omega t}$, but, in complex notation, we will omit the real part in the following.

The system is connected to a bath and it is described by a reduced density matrix $\rho$. The density matrix evolves under the master equation~\cite{Trivedi1988,Reulet1994}:
\begin{equation}\label{master2}
\frac{\pd \rho(t)}{\pd t}+\frac{i}{\hbar}[H(t),\rho(t)]
=-\gamma[\rho(t)-\rho_{\rm qe}(t)].
\end{equation}

Under the effect of the flux, a current $I$ is induced in the ring and is determined by the reduced density matrix
\begin{equation}\label{current}
I(t)=\tr[\rho(t) \mc J(t)]=I_0(\phi)+\delta I_\omega e^{-i\omega t}.
\end{equation}
The second equality represents the linear response of the current to the monochromatic excitation. The current operator $\mc J(t)$ in linear response reads
\begin{equation}
\mc J(t)=-\frac{\delta H(t)}{\delta\Phi(t)}
\simeq J-X\delta\phi_\omega e^{-i\omega t}\quad
\end{equation}
Remark that for the helical model in Eq.~(1) (in the main text) the current operator becomes time independent and equal to the static current operator $J=-ev_F/L\sigma_3$, and $X=0$. In contrast, the lattice model has a nonlinear flux dependence, which implies nontrivial expressions of $J$ and $X$. In linear response, we also expand the density matrix around the static value
\begin{equation}
\rho(t)=\rho_0+\delta\rho(\omega) e^{-i\omega t}.
\end{equation}
It is advantageous to work in the basis of the static Hamiltonian $H_0(\phi)$, $\{|n\ra \}$. Therefore the density matrix for the unperturbed system is $\rho_0=\sum_n f_n|n\ra\la n|$ with the Fermi-Dirac function $f_n(\e_n)$ depending on the energies of $H_0$.

From the second equality in Eq.~(\ref{current}), we obtain the explicit form for $I_0(\phi)$ and $\delta I_\omega$
\begin{eqnarray}\label{trI}
I_0(\phi)&=&\sum_nJ_{nn}f_n,\notag\\
\delta I_\omega&=&\sum_nX_{nn} f_n\delta\phi_\omega
+\sum_{mn}J_{mn}\delta\rho_{nm}(\omega),
\end{eqnarray}
where we have evaluated the operators in the basis of the static Hamiltonian.

Finally, the susceptibility is defined as the variation of the average current with respect to the flux variation:
\begin{equation}\label{chiGen}
\chi(\omega)=\frac{\delta I_\omega}{\delta\phi_\omega}.
\end{equation}
The static susceptibility follows by taking the zero frequency limit.

The last quantities to be determined are the components of the density matrix by solving the master equation in the eigenstate basis of $H_0$.
The master equation is represented as a set of differential equation. To linear order in $\delta\phi_\omega$, the equations for the matrix elements of the density operator are decoupled,  
\begin{eqnarray}\label{dens}
\delta\rho_{nn}(\omega)&=&-\frac{\pd f_n}{\pd\e_n}\frac{i\gamma_{nn}}{i\gamma_{nn}+\omega}(J_{nn}\delta\phi_\omega+\delta\mu_\omega)\notag\\
\delta\rho_{mn}(\omega)&=&-\frac{f_m-f_n}{\hbar\omega_{mn}}
\frac{\omega_{mn}-i\gamma_{mn}}{\omega_{mn}-\omega-i\gamma_{mn}}
J_{mn}\delta\phi_\omega,
\end{eqnarray}
where in the last equation $m\ne n$. The level separations were denoted: $\hbar\omega_{mn}=\hbar(\omega_m-\omega_n)$ and the chemical potential was also expanded near the static value $\mu=\mu_0+\delta\mu_\omega e^{-i\omega t}$. The condition that the number of particles is fixed reads as $\tr[\delta\rho(\omega)]=0$. This determines the change in the chemical potential with the flux:
\begin{equation}\label{variation}
\frac{\delta\mu_\omega}{\delta\phi_\omega}=
-\frac{\sum_n\frac{\pd f_n}{\pd\e_n}J_{nn}}
{\sum_n\frac{\pd f_n}{\pd\e_n}}.
\end{equation}
We will examine at the end of the section conditions for neglecting this term.

To first order in $\delta\phi_\omega$, the change in the current operator reads $\delta\mc J(t)=X\delta\phi(t)$.
The induced current in linear response is obtained using the equations for the density matrix~(\ref{dens}) with the current variation $\delta \mc J$ in Eq.~(\ref{trI}). Finally, Eq.~(\ref{chiGen}) yields a susceptibility that has diagonal and nondiagonal elements in the state basis of $H_0$,
\begin{eqnarray}\label{chiNum}
\chi(\omega)&=&\sum_{n} X_{nn}f_n
-J_{nn}(J_{nn}+\frac{\delta\mu_\omega}{\delta\phi_\omega})
\frac{\pd f_n}{\pd\e_n}
\frac{i\gamma_{nn}}{i\gamma_{nn}+\omega}\notag\\
&&-\sideset{}{'}\sum_{m,n}|J_{mn}|^2\frac{f_m-f_n}{\hbar\omega_{mn}}
\frac{\omega_{mn}-i\gamma_{mn}}{\omega_{mn}-\omega-i\gamma_{mn}}.
\end{eqnarray}
The primed sum denote in the following that $m\ne n$.
This formula is especially useful in the numerical determination of the susceptibility as it does not depend on the flux discretization.

The above equation is simplified using a sum rule from equating the second order perturbation theory for the eigenvalues $\e_n(\phi)$ of the Hamiltonian $H(t)$ and the Taylor expansion for the energy $\e_n(\phi+\delta\phi(t))$. Consequently, the static current and susceptibility operators are expressed as
\begin{equation}\label{exp}
J_{nn}=-\frac{\pd\e_n}{\pd\phi}=i_n,\quad
X_{nn}=\frac{\pd i_n}{\pd\phi}-2\sum_{m\ne n}\frac{|J_{mn}|^2}{\hbar\omega_{mn}}.
\end{equation}
The explicit formula for the tripartite susceptibility, $\chi=\chi_{\rm per}+\chi_D+\chi_{ND}$, follows using the sum rule in Eq.~(\ref{chiNum}): 
\begin{eqnarray}\label{chitot}
\chi(\omega)&=&\sum_n\frac{\pd(i_nf_n)}{\pd\phi}
-J_{nn}(J_{nn}+\frac{\delta\mu_\omega}{\delta\phi_\omega})
\frac{\pd f_n}{\pd\e_n}
\frac{i\omega}{\gamma_{nn}-i\omega}\notag\\
&&-\sideset{}{'}\sum_{m,n}|J_{mn}|^2\frac{f_m-f_n}{\hbar\omega_{mn}}
\frac{i\omega}{i(\omega_{mn}-\omega)+\gamma_{mn}}.\notag\\
\end{eqnarray}
The first term represents the persistent current contribution $\chi_{\rm per}$, while the second and the third terms stand, respectively, for the complex diagonal and nondiagonal susceptibilities.

The dissipative response for the system follows readily,
\begin{eqnarray}
\chi''(\omega)&=&
-\sum_nJ_{nn}(J_{nn}+\frac{\delta\mu_\omega}{\delta\phi_\omega})
\frac{\partial f_n}{\partial\epsilon_n}\frac{\omega\gamma_D}{\omega^2+\gamma_D^2}\notag\\
&&-\sideset{}{'}\sum_{m,n}|J_{mn}|^2\frac{f_m-f_n}{\hbar\omega_{mn}}
\frac{\omega\gamma_{mn}}{(\omega_{mn}-\omega)^2+\gamma_{mn}^2}.\notag\\
\end{eqnarray}
If one considers only the dynamics of populations and neglects the coupling to coherences (secular approximation, formally neglecting $\gamma_{mn}$ for $m\ne n$), or if one takes vanishing nondiagonal current components ($J_{mn}=0$), for a finite number of discrete levels, and a negligible variation in the chemical potential, then the susceptibility reads as
\begin{equation}
\chi=\sum_n\frac{\pd i_n}{\pd\phi}f_n-i_n^2\frac{\gamma_D}{\gamma_D-i\omega}
\frac{\pd f_n}{\pd\e_n}.
\end{equation}
This is the same expression which can be obtained using the simpler time-relaxation approximation
\begin{equation}
\frac{\pd\rho_{nn}}{\pd t}=-\frac{1}{\tau_n}(\rho_{nn}-f_n),
\end{equation}
with $\tau_n=\gamma_{nn}^{-1}$.

Lastly, let us return to the issue of the chemical potential variation $\delta\mu_\omega$. The condition that the number of particle is fixed for the time-independent problem $\sum_nf_n=N$, yields a constraint on the static chemical potential $\mu_0$. It follows from Eqs.~(\ref{variation}) and~(\ref{exp}) that in the linear response theory $\delta\mu_\omega/\delta\phi_\omega=\pd\mu_0/\pd\phi$. Therefore, a constant chemical potential with respect to the static flux, will have no time variations. This will prove important in the next section in the case of the helical model.

The chemical potential variation will equally prove negligible in the half-filling BHZ from a different point of view. Its contribution to the diagonal susceptibility is small in comparison to the other terms in Eq.~(\ref{chiNum}).
For example, after factoring out the dynamical dependence, the susceptibility $\chi_{\delta\mu}$ due to the variation of $\mu$ reads as
\begin{equation}\label{chimu}
\chi_{\delta\mu}\propto\frac{\big(\sum_n\frac{\pd f_n}{\pd\e_n}J_{nn}\big)^2}{\sum_m\frac{\pd f_n}{\pd \e_n}}\ll \sum_n J_{nn}^2\frac{\pd f_n}{\pd\e_n}.
\end{equation}
This is readily understood in the topological insulator case with helical edge states in the gap. The terms in the above sum contain mainly the contribution from the edge states, which are close to the Fermi energy. But the helical states have a linear energy-flux dispersion, and any current $J_{nn}$ has a partner with the same magnitude, but different sign, at a given energy. Then the numerators on the left hand side of the inequality give a vanishing contribution. For a finite temperature the denominator is finite and hence the response $\chi_{\delta\mu}$ is negligible. On the right hand side all the currents are squared, leading to a large contribution as observed in the body of the article.

\section{Helical models}
\subsection{Dirac ring}
The Dirac ring is described by the Hamiltonian $H=H_0+H'(t)$
\begin{equation}\label{hel2}
H_0=\frac{h v_F}{L}\bigg(-i\frac{\pd}{\pd\theta}+\frac{\phi}{\phi_0}
\bigg)\sigma_3,\quad H'(t)=\frac{ev_F}{L}\sigma_3\delta\phi(t).
\end{equation}
In this case, there is an infinite number of discrete eigenvalues with a linear dispersion, $\e_{n\sigma}=\sigma(n+\phi/\phi_0)$, where $\sigma$ indicates the spin degree of freedom $\sigma=\pm$ for $\up$, respectively $\down$. The spin component of the wave function are the eigenstates of $\sigma_3$ operator. The wave functions read as
\begin{equation}
|n\up\ra=e^{in\theta-i\e_{n\up}t}
\begin{pmatrix}
1\\0
\end{pmatrix},\quad |n\down\ra=e^{in\theta-i\e_{n\down}t}
\begin{pmatrix}
0\\1
\end{pmatrix}
.
\end{equation} 
Then the current operator matrix reads as
\begin{equation}
i_{n\sigma}=\avg{n\sigma|J|n'\sigma'}=-\sigma I_0\delta_{nn'}
\delta_{\sigma\sigma'},\quad I_0=\frac{ev_F}{L}.
\end{equation}

The model contains an infinite number of occupied states, which contribute to the persistent current. Thus the persistent current may not be a convergent sum, and subsequently $\chi_{\rm per}=\pd I_{\rm per}/\pd\phi$ may not be defined.
The authors have explicitly obtained in Ref.~\cite{Sticlet2013} the persistent current in the helical model using a regularization of the sum over the infinite number of states. The final result is finite and matches lattice results,
\begin{equation}\label{persistentCurrent}
\frac{I_{\rm per}}{I_0}=\sum_{m=1}^\infty\frac{2T/T^*}{\pi\sinh(mT/T^*)}
\sin(2\pi m\frac{\phi}{\phi_0})
\cos(2\pi m\frac{\mu}{\hbar\omega_0}).
\end{equation}
This regularization renders the persistent susceptibility well defined.
In contrast, the diagonal part of the susceptibility $\chi_D$ is always well defined as it contains predominantly the contribution from states near the Fermi surface due to the term $\pd f_{n\sigma}/\pd\e_{n\sigma}$.

Note that this current expression was obtained in the grand canonical ensemble for constant $\mu$.
Nevertheless, it can be connected with the case where the number of particles in the system is fixed. When $\mu=n\hbar\omega_0/2$ with $n$ integer, $\mu$ does not depend on the static flux and stands for a fixed number of particles. Because of the symmetry of the energy states, the $\mu=0$ case represents the half-filling case in the lattice models. Changing the number of particle by $n$ equivalent to a change in the chemical potential $\Delta\mu=\frac{n\hbar\omega_0}{2}$. Consequently, the change in the fermion parity leads to a shift by $n\phi_0/2$ in the current-flux characteristic. Due to gauge invariance, all physical quantities are periodic in $\phi_0$. Then adding an even number of particles is equivalent to the starting situation.

Because the chemical potential is constant in flux, the diagonal susceptibility reads
\begin{equation}
\chi_D=-\sum_{n\sigma}i_{n\sigma}^2
\frac{\pd f_{n\sigma}}{\pd\e_{n\sigma}}
\frac{i\omega}{\gamma_{nn}-i\omega},
\end{equation}
where $\gamma$ is spin independent.

The diagonal susceptibility $\chi_D$ can be expressed entirely in terms of the persistent current susceptibility $\chi_{\rm per}$.
After summing the spin degrees and algebraic manipulation of the sums in $\chi_D$, it follows that the diagonal susceptibility reads as
\begin{equation}\label{chihel}
\frac{\chi_D}{\chi_0}=\frac{i\omega}{\gamma_{nn}-i\omega}
\big(2+\frac{\chi_{\rm per}}{\chi_0}\big).
\end{equation}
The equation is then used to obtain the total susceptibility $\chi=\chi_{\rm per}+\chi_D$.
The dissipative susceptibility showed in the main body of the article follows in the approximation of identical $\gamma_{nn}=\gamma_D$ by taking the complex part of Eq.~(\ref{chihel}):
\begin{eqnarray}\label{chidiss2}
\frac{\chi_D''(\omega)}{\chi_0}&=&\frac{4\omega\gamma_D}{\omega^2+\gamma_D^2}
\bigg[\frac{1}{2}+\sum_{m=1}^\infty\frac{mT/T^*}{\sinh(mT/T^*)}\notag\\
&&\times\cos(2\pi m\frac{\phi}{\phi_0})
\cos(2\pi m\frac{\mu}{\hbar\omega_0})\bigg].
\end{eqnarray}

\subsection{Gapped Dirac ring}
The helical edge states are gapped in zero flux by adding a constant term which mixes the spin. This produces a vanishing diagonal susceptibility in zero flux.

The static Hamiltonian reads as 
\begin{equation}\label{gapped}
H_0=\hbar\omega_0
\bigg(-i\frac{\pd}{\pd\theta}+\frac{\phi}{\phi_0}\bigg)\sigma_3
+V\sigma_1,
\end{equation}
with the energy $\pm\e_n$,
\begin{equation}
\e_n=\big[\hbar^2\omega^2_0(n+\phi/\phi_0)^2+V^2
\big]^{1/2}.
\end{equation}

Let us consider again the same perturbation $H'(t)$, containing the time-oscillating flux.  
There are no current operator matrix elements between states with different angular momentum. Nevertheless, there are matrix elements between different spins.

The diagonal and nondiagonal current matrix elements are
\begin{equation}
|J_{nn}|=I_0\frac{\e_n(m=0)}{\e_n},\quad
|J_{n\bar n}|=I_0\frac{V}{\e_n},
\end{equation}
obeying the conservation law
$J_{nn}^2+J_{n\bar n}^2=I_0^2$. We have denoted here $\la n\!\up\!|J|n\!\down\ra=J_{n\bar n}$.

Without loss of generality, the chemical potential is taken at zero, $\mu=0$. The diagonal and nondiagonal susceptibility follow readily; the diagonal part reads as
\begin{equation}
\chi_D=\frac{i\omega}{\gamma_D-i\omega}\frac{\pi^2 T^*}{T}
\sum_n \frac{i_n^2}{\hbar\omega_0}\cosh^{-2}\big(\frac{\pi^2 T^*}{T}\frac{\e_n}{\hbar\omega_0}\big).
\end{equation}
The sum runs over all angular momenta $n$. Due to the fast decaying hyperbolic cosine at large $n$ the sum is quickly converging.

The nondiagonal susceptibility after summing over the spin degree of freedom reads as
\begin{equation}
\chi_{ND}=-
\sum_n\frac{J_{n\bar n}^2}{\e_n}
\tanh\big(\frac{\pi^2T^*}{T}\frac{\e_n}{\hbar\omega_0}\big)
\frac{\omega(\omega+i\gamma_{mn})}{(\omega+i\gamma_{mn})^2-4\frac{\e_n^2}{\hbar^2}}.
\end{equation}
The dissipative susceptibilities are obtained by taking the imaginary part in the above equations. These results allow direct comparison with dissipative susceptibilities in the BHZ model at half filling.

\section{Application to an ideal BHZ model}
To test the pertinence of using the helical model to deduce properties for edge states in a topological insulator, we consider Bernevig-Hughes-Zhang (BHZ) model~\cite{Bernevig2006} on a square lattice. In numerical simulations, we determine the linear response of the system to the time-dependent flux $\Phi(t)$ and show that it reproduces quite well the analytical results from the helical models.

The susceptibility in linear response is entirely determined by the lattice geometry, static Hamiltonian, static flux $\phi$, driving frequency $\omega$, damping rates $\gamma_{mn}$, temperature $T$, and number of particles $N$.
The susceptibility is determined in simulations using the linear response Eq.~(\ref{chiNum}). The following subsection will discuss the model, its ingredients, and  the various numerical tests used to extract the dissipative susceptibility.

\subsection{Model}
The BHZ model is implemented on a square lattice.
The Hamiltonian for the infinite system reads as
\begin{equation}
{\bf H}=\sum_{\mb k}c_{\mb k}^\dag\mc H(\mb k)c^{\phantom\dag}_{\mb k},
\end{equation}
where the spin indices for the creation and annihilation operators are implied from the structure of the first-quantized Hamiltonian,
\begin{equation}\label{matham}
\mc H(\mb k)=\begin{pmatrix}
h(\mb k) & 0 \\
0 & h^*(\mb -\mb k)
\end{pmatrix}
\end{equation}
with
\begin{eqnarray}
h(\mb k)&=&A[\sin(k_x)\sigma_1+\sin(k_y)\sigma_2]\notag\\
&&+\big[M-2B(2-\cos k_x-\cos k_y )\big]\sigma_3.
\end{eqnarray}
The coefficients $A$, $B$, and $M$ are material dependent parameters, which are taken in the simulation without reference to their exact values for the HgTe/CdTe quantum wells. Nonetheless, the parameters must obey a set of constraints in order for the system to be in a topological phase, in which edge states are localized near the two bases of the two bases of the BHZ cylinder: $A\ne 0$ and $M/B\in(0,8)$. In numerical simulations, the common choice was: $A=M=1$ and $B=0.6$.

\begin{figure}[t]
\includegraphics[width=0.45\columnwidth]{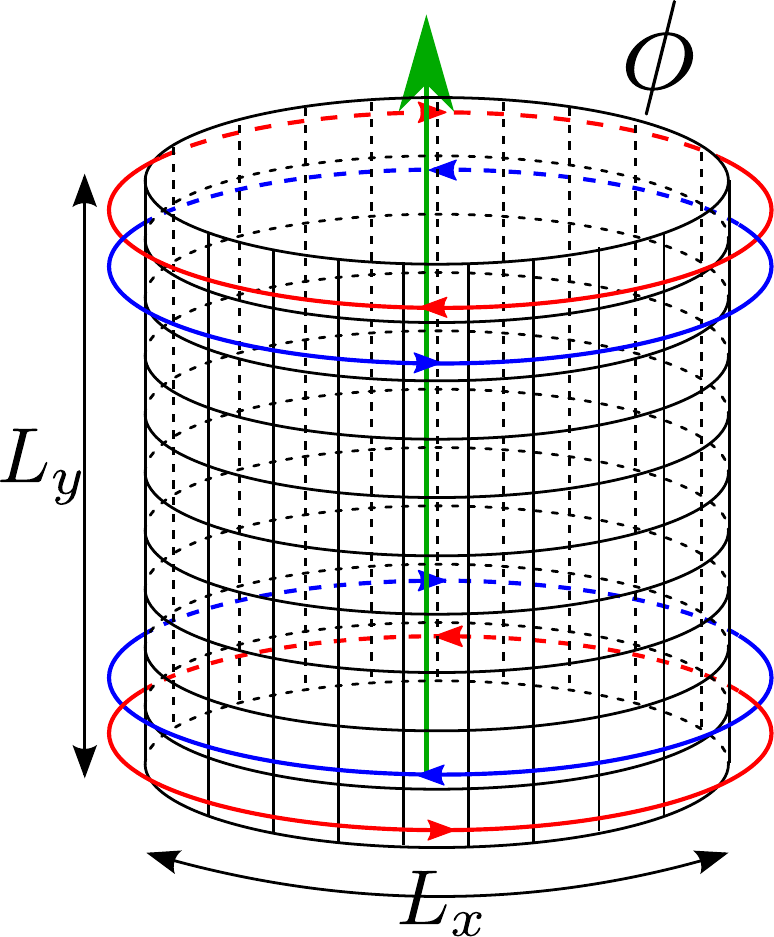}
\includegraphics[width=0.53\columnwidth]{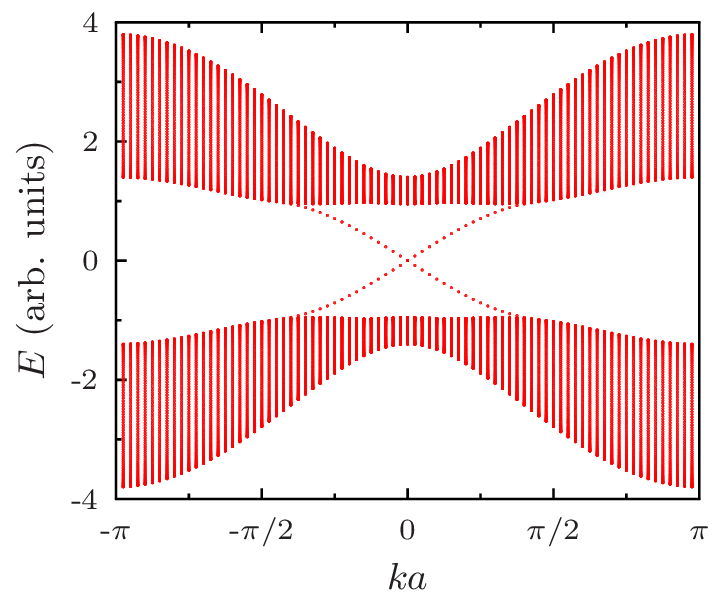}
\caption{(Color online). Right: The BHZ model in a cylindrical geometry. The cylinder has base circumference $L_x$ and height $L_y$. Left: Helical edge states with opposite spin form near the two bases of the cylinder. An oscillating flux $\phi$ threads the empty cylinder. The usual parameters used in simulations are $A=1$, $B=0.6$, $M=1$, ensuring that the system is in a topological phase. Right: The zero flux spectrum in a lattice of size $(L_x,L_y)=(80,80)a$.}
\label{fig:cyl}
\end{figure}

A finite square patch is cut out along the primitive lattice vectors from the infinite system and it is fashioned into a hallow cylinder. In the cylinder geometry, the coordinate $x$ counts the sites along the base of the cylinder, while $y$ counts the sites along the height of the cylinder (Fig.~\ref{fig:cyl}). Due to translational invariance in $x$ direction, the momentum $k$ parallel to the base is a good quantum number. Therefore, in a mixed representation, states can be described by momentum $k$, and real space, height index $y$.

There is a static flux $\phi$ threading the hollow cylinder. This is implemented in the lattice model through the Peierls substitution: 
\begin{equation}
k\to k+\frac{2\pi a}{L_x}\frac{\phi}{\phi_0}.
\end{equation} 
Therefore the current operator for the system threaded by the flux reads
\begin{eqnarray}\label{bhzcurrent}
J&=&-I_0\sum_{ky}
c_{ky}^\dag\bigg[\cos\big(k+\frac{2\pi\phi}{L_x\phi_0}\big)\sigma_1\tau_3\notag\\
&&-2\frac{B}{A}\sin\big(k+\frac{2\pi\phi}{L_x\phi_0}\big)\sigma_3\tau_0\bigg]c_{ky}.
\end{eqnarray}
Here, we have introduced another spin Pauli matrix $\tau$ relating the two blocks in the Hamiltonian~(\ref{matham}). Henceforth, the lattice constant is taken to be one, $a=1$ such that lengths $L_{x/y}$ can count the sites in $x/y$ direction.

The static Hamiltonian is diagonalized and one has access to its $4L_xL_y$ eigenstates $\e_n$ and eigenvectors $\{|n\ra\}$. It is apparent that the BHZ model has a chiral symmetry reflecting the property that each positive energy state has a partner at negative energy. Moreover, any state is at least twofold spin degenerate in zero flux.

For the given parameters, $A=1$, $B=0.6$, and $M=1$, the model is in a topological insulating phase with a bulk gap $\simeq 2A$. The edge states connect the bulk bands and traverse the gap. They are states living in the energy bulk gap and having a linear dispersion relation in momentum $k$ and in static flux $\phi$. 
The flux removes their spin degeneracy except at a set of flux values where the system recovers time-reversal invariance, $\phi=n\phi_0/2$, with $n$ any integer. Additionally, there is a degeneracy due to the fact that there are two edges, each accommodating a pair of edge states. Because the level-current amplitude is constant, the assumption that the diagonal rates are identical $\gamma_{nn}=\gamma_D$ is in effect. In contrast, the bulk states do not vary with the flux $\phi$, and therefore quantities that depend on them will be almost constant in flux.

\begin{figure}[t]
\centering
\includegraphics[width=0.49\columnwidth]{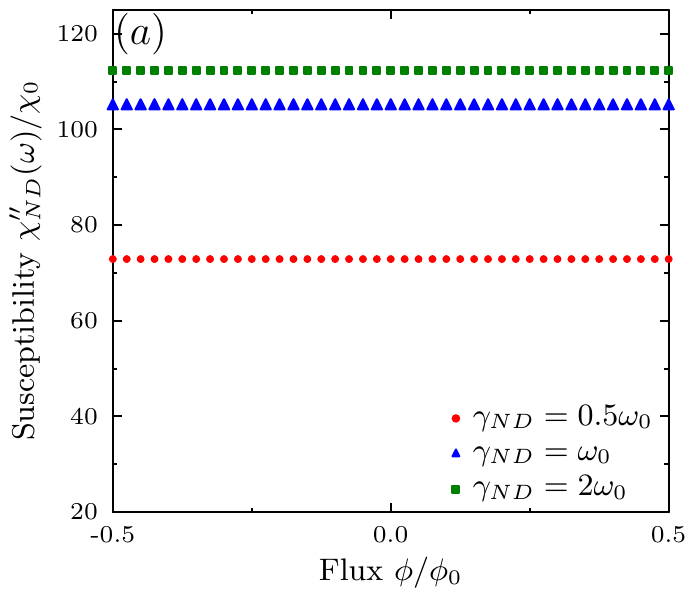}
\includegraphics[width=0.49\columnwidth]{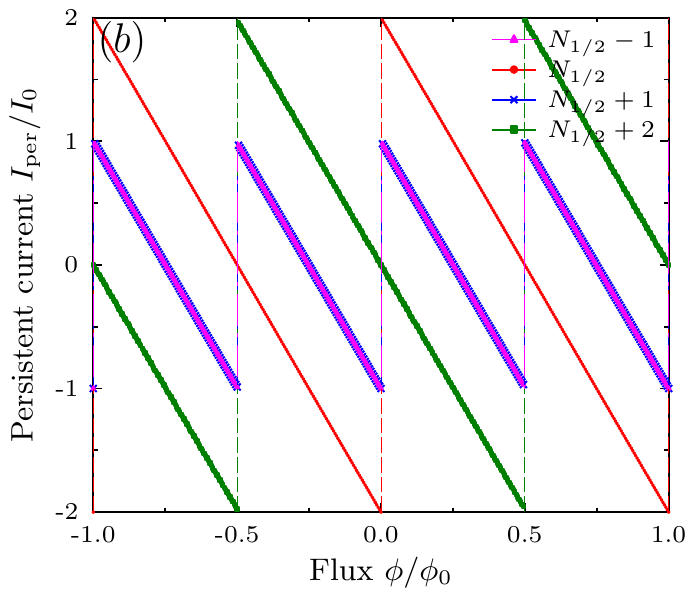}
\includegraphics[width=0.49\columnwidth]{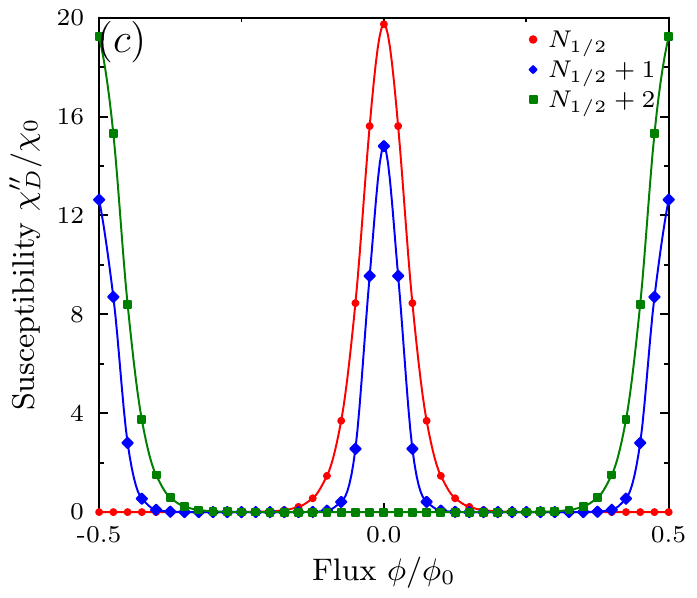}
\caption{(Color online). Different measurements for the dissipative susceptibilities $\chi_D''$ and $\chi_{ND}''$. (a) Trivial dependence of the nondiagonal susceptibility $\chi''_{ND}$ with the flux. Modulo 4 dependence on the particle number $N$ for the current (b) and susceptibility (c). (b) The persistent current is compared in lattice and helical number, at half filling $N_{1/2}=2L_xL_y$. (c) The diagonal susceptibility for three representative cases: half filling $N_{1/2}$, $N_{1/2}+2$ with a shift in the characteristic peak, and odd case $N_{1/2}+1$, with the peak split at both $\phi=0$ and $\pm\phi_0/2$. The lines in panel (c) are just guiding lines connecting the simulation points ($(L_x,L_y)=(8,60)a$).
Parameters of the system (unless otherwise stated): temperature $T=0.5T^*$,  frequency $\omega=\omega_0$, and lattice size $(L_x,L_y)=(80,80)a$. In the BHZ model $A=1=M$ and $B=0.6$.}
\label{fig:fsize}
\end{figure}

In the following, we work at (or close to) half filling, deep in the bulk gap. The dissipative susceptibilities in the model at half filling can be directly computed.
The current operator matrix elements are available. In contrast to the helical model, there are nondiagonal components, $J_{mn}\ne 0$ ($m\ne n$)~(\ref{bhzcurrent}). At half filling, the chemical potential does not depend on the flux $\mu=0$ and it allows us to introduce temperature in the model only through the equilibrium Fermi-Dirac functions $f_n=f(\e_n(\phi))$. The dependence of the chemical potential on the flux at different fillings will be discussed bellow.

Subsequently, the susceptibility in the model is computed using Eq.~(\ref{chiNum}). The bulk states contribute little to the diagonal susceptibility since  $\chi_D$ depends on states near the Fermi surface. At half filling, only the edge states are energetically close to $\mu=0$, and they yield the characteristic peak the dissipative susceptibility (see main body of the article). The edge states do not contribute to the nondiagonal susceptibility $\chi_{ND}$ since the driving frequency cannot induce spin flips or changes in angular momentum. In contrast, since the current has off-diagonal components between the bulk edge states, they yield a large paramagnetic contribution to the nondiagonal dissipative susceptibility. However, for large systems, the bulk states and the nondiagonal susceptibility depend little on the flux  (see Fig.~\ref{fig:fsize}(a)). Therefore the characteristic peak in the dissipative susceptibility contains information only from the edge states. The peak is maximal when the driving frequency is equal to the diagonal dissipation rate $\gamma_D$. This allows in turn to determine the lifetime of the edge states $\gamma_D^{-1}$.

\subsection{Parity effects in the lattice model}
The total number of available states in the lattice model is $N=4L_xL_y$. In the main body of the article, we have worked at half filling $N_{1/2}=2L_xL_y$, where the number of particles is even. In the present section, we discuss effects due to changes from half filling, while still remaining in the bulk gap.

The particle number enters into the equation through the chemical potential, in the Fermi-Dirac function
\begin{equation}
f_n=\frac{1}{e^{\beta(\e_n-\mu)}+1}.
\end{equation}
The condition that the number of particles is fixed imposes constraints on the chemical potential. In particular, $\mu$ is determined from the normalization condition $N=\text{Tr}[f_n]$, with the trace over all the eigenstates. Thus the chemical potential is generally a function of the flux and the number of particles $N$. Nevertheless for certain constant values, $\mu$ does not vary with the number of particles. When $\mu=n\hbar\omega_0/2$, with $n$ any integer, there is always an even number of particles $N$ in the model. Indeed, since all the energy states are at least twofold degenerate, fixing the chemical potential at $\mu=n\hbar\omega_0/2$ allows one to scan the ground state in an entire period in the energy-flux dispersion, while conserving the particle number.
In contrast, a constant $\mu$ cannot capture the cases with odd number of particles in the lattice.

In case of even $N$, when adding or subtracting $2n$ particles at half filling, the current-flux and susceptibility-flux characteristics will shift by half flux quantum $\phi_0/2$ for $n$ odd, and due to gauge invariance, they will be identical for $n$ even.

In case of odd $N$, the situation is more complicated, with a flux-dependent chemical potential.
In contrast with the helical model or the BHZ even filling, there is an additional term in the diagonal susceptibility, ensuring the conservation of particle number,
\begin{equation}
\chi_D=-\sum_{n}
J_{nn}(J_{nn}+\frac{\pd\mu_0}{\pd\phi})
\frac{\pd f_{n}}{\pd e_n}
\frac{i\omega}{\gamma_{nn}-i\omega}.
\end{equation}
The chemical potential is obtained by inverting numerically the relation $\tr[f_n]=N$. All odd particle cases are distinguished by peaks in the susceptibility both at $\phi=0$ and $\phi=\phi_0/2$. These peaks are smaller in amplitude in comparison with the even cases. 

The upshot of the section is that there is a dependence modulo 4 on the number of particles in the model. For even number of particles there will be a shifts from $\phi/\phi_0=0$ to $0.5$ of the susceptibility signal. For odd number of particles, the signal is split between peaks at both time-invariant flux values. 


The helical model can be used to provide more understanding to these parity effects. It can account and explain the particular features in the response.
\subsection{Comparison with the helical model}
Before studying the dynamical response of the system, let us compare the helical and the BHZ models at zero temperature in the absence of driving. The interesting physics in this case is that of equilibrium persistent currents.

In the BHZ model there are one pair of helical states  at both bases of the cylinder. Hence one has to employ two helical models to account for the lattice model. Moreover, in order to compare the models, it is necessary to scale the physical quantities according to the appropriate level spacing. 
Energies in the helical model are scaled with $\hbar\omega_0$, while in the BHZ model, they are scaled with the level spacing at the Fermi surface, $\frac{2\pi a}{L_x}A$, with $a$ the lattice constant and $A$ a BHZ model parameter. Similarly, all the other characteristic quantities, $I_0$ and $\chi_0$, are related between the two models. Finally, we work in units where the lattice spacing is dimensionless $a=1$ and $\hbar=1$.

In the numerical simulation for the BHZ model, we obtain a persistent current which close to half filling $N_{1/2}=2L_xL_y$ depends modulo 4 on the number of particles. This is represented for relevant cases in Fig.~\ref{fig:fsize}(c), $N_{1/2}$, $N_{1/2}\pm 1$ and $N_{1/2}+2$. This can be understood by tracking the particles near zero energy. At half filling $N_{1/2}$ there are two filled states and two empty states at zero energy and zero flux. The $4I_0$ discontinuity in the current at zero flux indicates that the ground state at negative and positive flux is quite different. At small negative flux there is an imbalance, two filled right-moving states and two empty left-moving states, while at positive flux it is the reverse. If all the states are filled in zero-flux, one encounters the same difference in the ground-state moved at half-integer flux $\phi/\phi_0=\pm 1/2$. In other words, at $N_{1/2}+2$, the current-flux characteristic has shifted by $\phi_0/2$. At odd number of particles $N_{1/2}\pm 1$, discontinuities appear at integer and half-integer flux, but the amplitude was halved. This is because the current carried by the almost fourfold degenerate states near $\phi=0$ and $\phi=\phi_0/2$ is always only $\pm I_0$.

The helical model can account perfectly for the persistent currents in the BHZ model. Let us denote the lattice persistent current as $I_{\rm per}^{\rm bhz}$. Two helical models are required to mimic the two pair of edge states in the BHZ model. Let us denote by $I_{\rm per}(\mu)$ the current for one helical model~(\ref{persistentCurrent}). As noted before, adding a particle is equivalent to varying the chemical potential in the helical model by half energy spacing $\Delta\mu=\hbar\omega_0/2$. Then the currents in the BHZ model are obtained by adding or subtracting particles in the two helical models. For the cases with an even number of particles, represented in the Fig.~\ref{fig:fsize}(c).
\begin{equation}\label{comp}
\frac{I^{\rm bhz}_{\rm per}(N_{1/2})}{I_0^{\rm bhz}}
=2\frac{I_{\rm per}(0)}{I_0},\quad
\frac{I^{\rm bhz}_{\rm per}(N_{1/2}+2)}{I_0^{\rm bhz}}
=2\frac{I_{\rm per}(\frac{\hbar\omega_0}{2})}{I_0}.
\end{equation}
Similarly, for the odd particle cases, one extra particle is added or extracted in one of the helical models
\begin{equation}
\frac{I^{\rm bhz}_{\rm per}(N_{1/2}\pm 1)}{I_0^{\rm bhz}}=
\frac{I_{\rm per}(0)
+I_{\rm per}(\pm\frac{\hbar\omega_0}{2})}{I_0}.
\end{equation}
Thus the helical models explain the modulo 4 pattern in the persistent current simulations.

\begin{figure}[t]
\includegraphics[width=0.60\columnwidth]{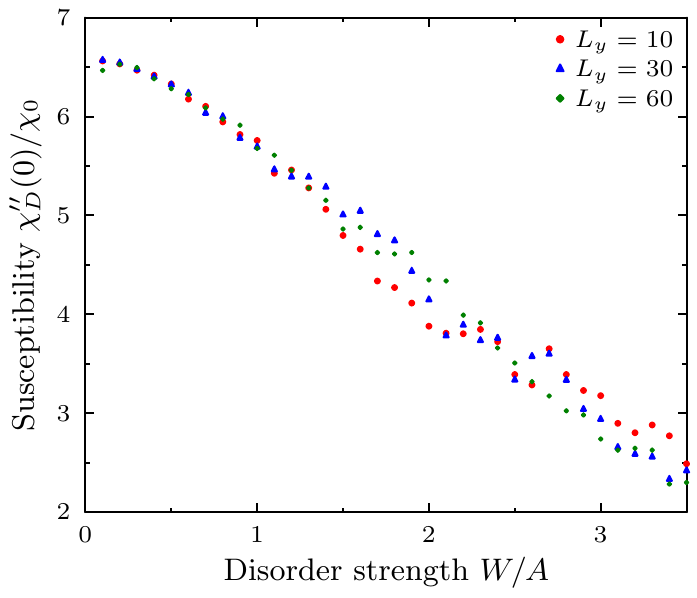}
\caption{(Color online). Diagonal dissipative susceptibility in zero flux as a function of disorder strength. The susceptibility is averaged over 200 realizations for three different cylinder heights $L_y$. The signal in zero flux decreases linearly with disorder. System parameters: $L_x=8a$, $(A,B,M)=(1,0.6,1)$, and $T=1.5T^*$.}
\label{fig:disorder}
\end{figure}

The above arguments hold qualitatively also at higher temperature in the presence of driving. For even particle cases, the signal in the susceptibility is correctly given by doubling the signal in the helical cases. For the odd particle numbers, the BHZ signal is not exactly given by the sum of two helical models  shifted by $\Delta\mu=\hbar\omega_0/2$. The direct sum of helical models would predict a signal split at both $\phi=0$ and $\phi=\phi_0/2$, and half the size
of the signal in the even case. In the simulation we see that indeed the signal is split, but its amplitude is $50\%$ higher than the predicted signal under the above simple argument (Fig.~\ref{fig:fsize}(d)).

Throughout the main body of the article, we have worked at half filling with the number of particles $N_{1/2}=2L_xL_y$.
In this case the chemical potential $\mu=0$ is constant as a function of the static flux.

\subsection{Scalar disorder}
To test the robustness of the signal to addition of disorder, the initial BHZ model under flux is enriched with scalar disorder on-site disorder
\begin{equation}
H''=w_j\sum_{j}c_j^\dag\sigma_0\tau_0c_j.
\end{equation}
The on-site disorder $w$ is a random variable, uniformly distributed in the interval $[-W/2,W/2]$, where $W$ is the disorder amplitude. In simulations, the disorder is taken in units of model parameter $A$.

As a proof of principle, we consider ideally thin cylinders $L_x=8a$. This allows exploring long cylinder lengths averaged over many disorder realizations and obtaining readily the diagonal dissipative response. 
The system shows sensitivity to disorder, and the value of the disorder average of the susceptibility decreases continuously with disorder. Nevertheless, the dissipative susceptibility never vanishes as in the case of scalar disorder in a nonrelativistic fermion systems. 

\begin{figure}[t]
\centering
\includegraphics[width=0.8\columnwidth]{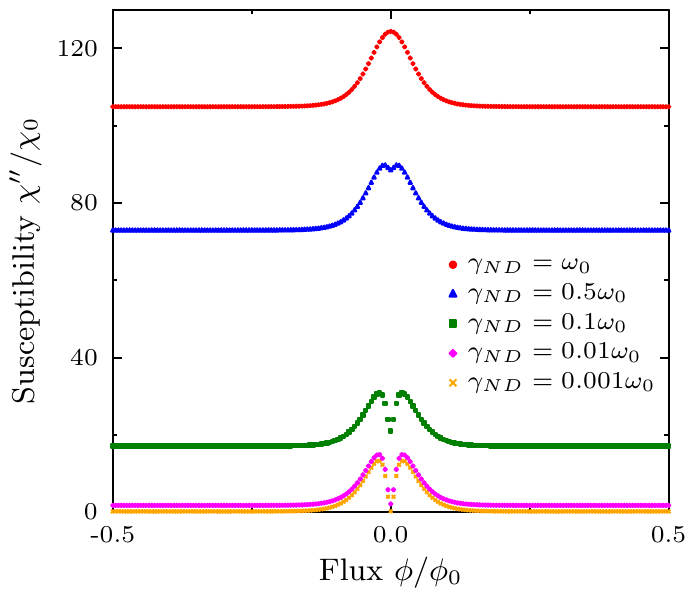}
\caption{(Color online). Total dissipative susceptibility $\chi''$ as a function of flux, for different values of the nondiagonal damping rates, in the presence of a small Zeeman energy $V=0.01\hbar\omega_0$, in the cylindrical BHZ model. The dissipative contribution from the bulk states increases with larger damping rates. The diagonal susceptibility contribution is washed out at large coherence damping rates $\gamma_{ND}$. The driving frequency is set to give the maximal diagonal susceptibility, $\omega=\gamma_{D}$ (and $\omega=\omega_0$).
System parameters are $(A,B,M)=(1,0.6,1)$ at half filling for a lattice size $(L_x,L_y)=(80,80)a$.
}
\label{fig:gndz}
\end{figure}

\subsection{Constant Zeeman field}
Finally, the BHZ model is subjected to a constant transversal Zeeman field which gaps the edge states in zero flux. The field mixes the spin states and leads to the observed decrease in the diagonal susceptibility at zero flux.

The perturbation added to the Hamiltonian is constant for all the sites in the lattice
\begin{equation}\label{Z}
H'''=V\sum_jc_j^\dag\sigma_0\tau_1c_j,
\end{equation}
where $j$ runs over all the sites in the cylinder and $V$ is a constant Zeeman energy.
The Zeeman term anticommutes with the BHZ Hamiltonian~(\ref{matham}) and mixes the spin states.
To compare the response in the helical and BHZ models, the field is scaled with the respective energy level spacing near the Fermi energy, $\hbar\omega_0$. This yields again agreement between the diagonal response in the two systems (Fig.~3 in the main body of the article).
Very small fields with respect to the bulk gap can still create infinitesimal gaps in the helical edge states at zero flux. This affects the states infinitesimally close to the zero flux by creating a vanishing diagonal susceptibility. This is reflected as a dip of very small width in the diagonal signal.

In the presence of a magnetic field the effect of large coherence damping rates $\gamma_{ND}$ can wash out the diagonal susceptibility signal. The peak at zero flux in the nondiagonal dissipative susceptibility, due to spin mixing of the edge channels, dominates the diagonal susceptibility dip predicted from the vanishing of the level current at the time-reversal invariant fluxes. Additionally, interband transitions between the bulk states are enhanced at larger damping rates $\gamma_{ND}$. The bulk states contribution remains almost constant in flux and it adds to uniformly increase the overall dissipative susceptibility (Fig.~\ref{fig:gndz}).
\end{document}